\documentclass[conference]{IEEEtran}
\IEEEoverridecommandlockouts
\usepackage{amsthm,amsmath,amssymb,mathtools,bm,etoolbox,float,hyperref}
\usepackage[dvipsnames]{xcolor}
\usepackage{graphicx,cite,cuted}
\usepackage{bigints}
\usepackage{caption}

\usepackage{stackengine}

\usepackage{balance}

\theoremstyle{corollary}
\newtheorem{corollary}{Corollary}

\usepackage{mathrsfs,verbatim}
\usepackage{ellipsis}
\usepackage{tikz}
\usepackage{tikz-3dplot}
\usepackage[utf8]{inputenc}
\usepackage{pgfplots} 
\usepackage{pgfgantt}
\usepackage{pdflscape}
\pgfplotsset{compat=newest} 
\pgfplotsset{plot coordinates/math parser=false}
\pgfplotsset{every  tick/.style={black,},ylabel style={font=\tiny},xlabel style={font=\tiny},tick label style={font=\tiny},legend style= {font=\scriptsize},
minor x tick num=1,minor y tick num=1,xminorticks=true,yminorticks=true,}
  \newlength\fheight
\newlength\fwidth

\usepackage{xinttools} % for the \xintFor***
\usepackage{tikz}
\usepackage{tkz-euclide}
\usetikzlibrary{calc}
\newtheorem{theorem}{Theorem}
\newtheorem{proposition}{Proposition}

\newtheorem{remark}{Remark}

\def\biglen{20cm} % playing role of infinity (should be < .25\maxdimen)
\tikzset{
  half plane/.style={ to path={
       ($(\tikztostart)!.5!(\tikztotarget)!#1!(\tikztotarget)!\biglen!90:(\tikztotarget)$)
    -- ($(\tikztostart)!.5!(\tikztotarget)!#1!(\tikztotarget)!\biglen!-90:(\tikztotarget)$)
    -- ([turn]0,2*\biglen) -- ([turn]0,2*\biglen) -- cycle}},
  half plane/.default={1pt}
}

\DeclareMathAlphabet{\pazocal}{OMS}{zplm}{m}{n}

\usepackage{caption}
\usepackage{multicol,subcaption}
\usepackage{balance}

\def\BibTeX{{\rm B\kern-.05em{\sc i\kern-.025em b}\kern-.08em
    T\kern-.1667em\lower.7ex\hbox{E}\kern-.125emX}}
\begin{document}

\title{Forward Link Analysis for Full-Duplex Cellular Networks with Low Resolution ADC/DAC}

\author{\IEEEauthorblockN{Elyes Balti and Brian L. Evans}\thanks{This work was supported by AT\&T Labs and NVIDIA, affiliates of the 6G@UT Research Center within the Wireless Networking and Communications Group at UT Austin.}
\IEEEauthorblockA{\textit{6G@UT Research Center} \\
\textit{Wireless Networking and Communications Group} \\
\textit{The University of Texas at Austin} \\
ebalti@utexas.edu, bevans@ece.utexas.edu}
}

\maketitle

\begin{abstract}
In this work, we consider a full-duplex (FD) massive multiple-input multiple-output (MIMO) cellular network with low resolution analog-to-digital converters (ADCs) and digital-to-analog converters (DACs). Our first contribution is to propose a unified framework for forward link analysis where matched filter precoders are applied at the FD base stations (BSs) under channel hardening. Second, we derive expressions for the signal-to-quantization-plus-interference-plus-noise ratio (SQINR) for general and special cases. Finally, we quantify effects of quantization error, pilot contamination, and full duplexing for a hexagonal cell lattice on spectral efficiency and cumulative distribution function (CDF) to show that FD outperforms half duplex (HD) in a wide variety of scenarios.
\end{abstract}


\section{Introduction}
Cellular networks are experiencing difficult design challenges to provide ever increasing data rates.
Two tried approaches over the last two decades have been to increase transmission bandwidths and spectral efficiency.

To overcome the limited available bandwidth and support a large number of users, massive multiple-input multiple-output (MIMO) has emerged as a solution to improve the data rate without increasing the bandwidth \cite{mimo2}. A massive number of antennas can also provide significant multiplexing gain and serve a large number of users simultaneously \cite{mimo3}. With a massive number of antennas comes significant challenges in power consumption, energy efficiency, channel modeling and estimation due to the increased dimensionality \cite{overview}.

% Using a massive number of antennas causes a corresponding increase in the channel dimension, and this paper helps address the need to accurate analytical channel modeling and solutions.
% including supporting a larger number of maximum users increasing number of active users per cell becomes hefty and the dimensionality of the channels grows very large. These difficulties call for the needs to develop  and such is the subject of this paper. 

To further improve efficiency of resource allocations, full-duplex (FD) operation has been proposed as a robust solution to not only double the spectral efficiency but also reduce latency \cite{ianMagazine}. In addition, FD systems are cost-efficient since transmit and receive arrays can be reduced to a single array wherein the transmission and reception are shared. These benefits call for the possible applicability of FD in practice, e.g. machine-to-machine communications and integrated access and backhaul which is suggested in 3GPP Release 17 \cite{zf,elyesiab,release17,R6,R7}. Due to the near-far problem, FD systems are vulnerable to loopback self-interference (SI), which leaks from the transmit arrays to the receive arrays at the FD transceiver and is several orders of magnitude stronger than the uplink (reverse link) received signal power. In cellular networks, the uplink users are corrupted by the SI; without proper SI reduction, the FD systems end up completely dysfunctional \cite{zfjournal,zf,ianTWC}. Since the FD BS simultaneously transmits to downlink users and receives from uplink users, the downlink users are vulnerable to the inter-user-interference (IUI) caused by the uplink users. FD systems directly affect uplink users by the loopback SI and indirectly degrades downlink (forward link) reception due to the IUI introduced by the uplink users \cite{massiveadc}.

The combination of FD with massive MIMO may circumvent these two limitations \cite{R6,R7,massiveadc,R9}. By increasing the number of antennas, enough degrees of freedom become available to suppress SI and IUI \cite{massiveadc,elyesjoint}. Although this increases spectral efficiency, employing a massive number of antennas may lead to hefty power consumption, esp. for all-digital systems wherein each antenna has an RF chain with a phase shifter and ADC/DAC \cite{overview}. To address this shortcoming, low-resolution ADCs/DACs have emerged as a viable solution to reduce the power consumption at the expense of sacrificing a portion of the achievable rate \cite{R3,R4,R5}. 

In this context, we propose the modeling of FD massive MIMO cellular networks with low-resolution ADCs/DACs and under pilot contamination. The BSs operate in FD mode while the users equipments (UEs) are equipped with single antennas and operate in full-resolution and half-duplex mode. Communication performance is simulated for hexagonal lattice with two tiers. To the best of our knowledge, this is the first work that considers the analysis of FD massive MIMO cellular networks with low resolution ADCs/DACs under pilot contamination for the forward link. The contribution of this work attempts to account for all the physical layer irregularities and imperfections in order to develop an accurate analytical modeling to address the many performance limitations. 

Section II discusses the network model and Section III analyzes the forward link. Numerical results are reported in Section IV while concluding remarks appear in Section V.

{\bf Notation}. Italic non-bold letters refer to scalars while bold lower and upper case stand for vector and matrix, respectively. We denote subscripts $u$ for uplink and $d$ for downlink.
\section{Network Model}
We consider a macrocellular network where each BS operates in FD mode and is equipped with $N_{\mathsf{a}} \gg 1$ antennas. Each UE operates in half-duplex mode and has a single antenna.
\subsection{Large-Scale Fading}
Each UE is associated with the BS from which it has the highest large-scale channel gain. We denote $K_\ell^u$ and $K_\ell^d$ to be the number of uplink and downlink UEs served by the $\ell$-th BS. The large-scale gain between the $\ell$-th BS and the $k$-th user connected to the $l$-th BS comprises pathloss with exponent $\eta > 2$ and independently and identically distributed (IID) shadowing across the paths, and is defined as
\begin{equation}
G_{\ell,(l,k)} = \frac{L_{\text{ref}}}{r_{\ell,(l,k)}^\eta} \chi_{\ell,(l,k)}  
\end{equation}
where $L_{\text{ref}}$ is the pathloss intercept at a unit distance, $r_{\ell,(l,k)}$ the link distance, and $\chi_{\ell,(l,k)}$ as the shadowing coefficient satisfying $\mathbb{E}[\chi^\delta] < \infty$, where $\delta = 2/\eta$. 

Without loss of generality, we denote the 0-th BS as the focus of interest and we drop its subscript.

Further, we introduce large-scale fading between the UEs to model the inter-user communication. We denote $T_{(\ell,n),(l,k)}$ as the large-scale channel gain between the $n$-th user and the $k$-th user associated with the $\ell$-th and $l$-th BSs, respectively.
\subsection{Small-Scale Fading}
We denote $\boldsymbol{h}_{\ell,(l,k)} \sim \mathcal{N}_{\mathbb{C}}(\bold{0},\boldsymbol{I})$ as the normalized reverse link $N_{\mathsf{a}} \times 1$ small-scale fading between the $k$-th user located in cell $l$ and the BS in cell $\ell$ and $\boldsymbol{h}^*_{\ell,(l,k)}$ as the forward link reciprocal, assuming time division duplexing (TDD) with perfect calibration \cite{massiveforward}. In addition, we denote by $g_{(\ell,n),(l,k)} \sim \mathcal{N}_{\mathbb{C}}(0,1)$ the $1 \times 1$ small-scale fading between the $n$-th user in cell $\ell$ and the $k$-th user in cell $l$ \cite{massiveadc}. Note that $(\cdot)^*$ is the Hermitian operator. We denote by $\boldsymbol{h}^*_{\ell,(0,k)}$ the channel between the $\ell$-th BS and $k$-th UE in the cell of interest. To the sake of notation, we drop the index 0 for the cell of interest and the fading becomes $\boldsymbol{h}_{\ell,k}$, while the fading the between the $k$-th UE and its served BS in the cell of interest is $\boldsymbol{h}_k$.
\subsection{Low Resolution ADC/DAC}
Without loss of generality, we analyze the low resolution ADC/DAC in single-cell single-user scenario. Then we generalize the analysis for forward link cellular network.
The downlink unquantized precoded signal $\boldsymbol{x}_d$ is given by
\begin{equation}
\boldsymbol{x} = \boldsymbol{F}\boldsymbol{s}     
\end{equation}
where $\boldsymbol{F}$ is the precoder. As the additive to quantization plus noise model (AQNM) approximates the quantization error \cite{R6,R7,R8}, the transmitted signal $\boldsymbol{x}_{q}$ after the DAC at the BS can be obtained by
\begin{equation}
   \boldsymbol{x}_{q} = \alpha \boldsymbol{x} + \boldsymbol{q}    
\end{equation}
where $\boldsymbol{q}$ is the AQNM, while $\alpha = 1-\rho$ and $\rho$ is the inverse of the signal-to-quantization-plus-noise ratio (SQNR), which is inversely proportional to the square of the resolution of an ADC, i.e., $\rho \propto 2^{-2b}$. Table \ref{etaparam} defines the values of $\rho$ with respect to the number of bits.
\begin{table}[b]
\renewcommand{\arraystretch}{1}
\caption{$\rho$ for different values of $b$ \cite{massiveadc}.}
\label{etaparam}
\centering
\begin{tabular}{cccccc}
$\boldsymbol{b}$ & 1 & 2 & 3 & 4 & 5\\
\hline
$\boldsymbol{\rho}$ & 0.3634 & 0.1175 & 0.03454 & 0.009497 & 0.002499
\end{tabular}
\end{table} 

We define the AQNM covariance matrix as follows \cite{massiveadc}
\begin{equation}
\boldsymbol{R}_{\boldsymbol{q}} = \mathbb{E}[\boldsymbol{q}\boldsymbol{q}^*] = \alpha(1-\alpha)\text{diag}\left(\boldsymbol{F}\boldsymbol{F}^* \right)
\end{equation}

\section{Performance Analysis}
Before the DAC, the signal transmitted by the $\ell$-th BS is
\begin{equation}
\boldsymbol{x}_{\ell} = \sum_{k=0}^{K_\ell^d-1} \sqrt{\frac{P_{\ell,k}}{N_{\mathsf{a}}}}\boldsymbol{f}_{\ell,k}s_{\ell,k}    
\end{equation}
where $P_{\ell,k}$ is the power allocated to the data symbol $s_{\ell,k} \sim \mathcal{N}_{\mathbb{C}}(0,1)$, which is precoded by $\boldsymbol{f}_{\ell,k}$ and intended for its $k$-th user. The power allocation satisfies 
\begin{equation}
    \sum_{k=0}^{K_\ell^d-1}P_{\ell,k} = P_d
\end{equation}
where $P_d$ is the power budget at the BS. Since the BS operates in FD mode, the uplink users are only corrupted by the SI while the downlink users are SI-free. But since the uplink users are sending simultaneously when the downlink users are receiving, the latter are vulnerable to the inter-user interference caused by the uplink users.
\subsection{Matched Filter Precoder}
From the reverse link pilots transmitted by its users,
the $\ell$-th BS gathers channel estimates $\hat{\boldsymbol{h}}_{\ell,(\ell,0)},\ldots,\hat{\boldsymbol{h}}_{\ell,(\ell,K_\ell^d-1)}$. With matched filter transmitter, the precoders at cell $\ell$ are given by
\begin{equation}
\boldsymbol{f}_{\ell,k}^{\mathsf{MF}} = \sqrt{N_{\mathsf{a}}}\frac{\hat{\boldsymbol{h}}_{\ell,(\ell,k)}}{\sqrt{\mathbb{E}\left[ \left\|  \hat{\boldsymbol{h}}_{\ell,(\ell,k)} \right\|^2 \right]}},~k=0,\ldots,K_\ell^d-1    
\end{equation}
Under pilot contamination, a matched filter for user $k$ satisfies $\boldsymbol{f}^{\mathsf{MF}}_k \propto \hat{\boldsymbol{h}}_k$. This entails the following expression as \cite[Eq.~(10.104)]{foundationsmimo} 
\begin{equation}
\begin{split}
\boldsymbol{f}^{\mathsf{MF}}_k =&\sqrt{\frac{\frac{P_k}{P_u}\mathsf{SNR}^u_k}{1+\frac{P_k}{P_u}\mathsf{SNR}^u_k + \sum_{\ell \in \mathcal{C}}\frac{P_{\ell,k}}{P_u}\mathsf{SNR}^u_{\ell,k} } }\\&\times\left( \boldsymbol{h}_k + \sum_{\ell \in \mathcal{C}} \sqrt{\frac{\frac{P_{\ell,k}}{P_u}\mathsf{SNR}^u_{\ell,k}}{\frac{P_k}{P_u}\mathsf{SNR}_k^u}}\boldsymbol{h}_{\ell,k} + \boldsymbol{v}^{'}_k \right)  
\end{split}
\end{equation}
Where $\mathcal{C}$ is the set of cells reusing the same pilot dimensions and $\mathsf{SNR}^u_{\ell,k}$ is the SNR of the link between the $k$-th uplink in the cell of interest and the $\ell$-th BS. Similarly, $\mathsf{SNR}$ is the $\mathsf{SNR}^u_k$ is the SNR of link between the $k$-th uplink UE and the its serving BS of interest. We define $\mathsf{SNR}_{\ell,k}^u = G_{\ell,k}P_u/N_0$, $\mathsf{SNR}_{\ell,k}^d = G_{\ell,k}P_d/N_0$, $N_0$ is the noise power and $P_u$ is the power budget at each user.

The scaling is important to operate the decoder, but immaterial since it equally affects the received signal as well as the noise and the interference. With scaling such that $\mathbb{E}\left[\| \boldsymbol{f}_k^{\mathsf{MF}} \|^2\right] = N_{\mathsf{a}}$ and with the entries of $\boldsymbol{v}_k^{\prime}$ having power $1/\left( \frac{P_k}{P_u}\mathsf{SNR}_k^u \right)$. The pilots are assumed to be regular and aligned at every cell.
\begin{corollary}\label{corchannelhardening}
The matched filter precoder $\boldsymbol{f}^{\mathsf{MF}}_k$ has the following properties
\begin{enumerate}
    \item $\mathbb{E}\left[\|\boldsymbol{f}^{\mathsf{MF}}_k \|^2 \right] = N_{\mathsf{a}}$.
    \item $\mathbb{E}\left[\| \boldsymbol{f}^{\mathsf{MF}}_k \|^4 \right] = N_{\mathsf{a}}^2 + N_{\mathsf{a}}$.
    \item $\left|\mathbb{E}\left[\boldsymbol{h}_k^*\boldsymbol{f}_k^{\mathsf{MF}}\right]\right| = N_{\mathsf{a}}$.
    \item $\mathbb{E}\left[ \left| \boldsymbol{h}^*_{\ell,k}\boldsymbol{f}^{\mathsf{MF}}_k \right|^2 \right] = N_{\mathsf{a}}$.
\end{enumerate}
\end{corollary}
\subsection{Channel Hardening}
Since we consider receivers reliant on channel hardening, the $k$-th user served by the BS of interest regards $\mathbb{E}[\boldsymbol{h}^*_{k}\boldsymbol{f}_k]$ as its precoded channel wherein the small-scale fading is averaged. The variation of the actual precoded channel around the mean incurs self-interference, such that the received signal can be formulated as (\ref{downlink2}).
\begin{figure*}
\begin{equation}\label{downlink2}
\begin{split}
    y_{q,k} =& \underbrace{\alpha \sqrt{\frac{G_{k}P_k}{N_{\mathsf{a}}}} \mathbb{E}[\boldsymbol{h}^*_{k}\boldsymbol{f}_k]s_k}_{\mathsf{Desired Signal}} + \underbrace{\alpha \sqrt{\frac{G_{k}P_k}{N_{\mathsf{a}}}}\left( \boldsymbol{h}^*_{k}\boldsymbol{ f}_k -  \mathbb{E}[\boldsymbol{h}^*_{k}\boldsymbol{f}_k] \right) s_k}_{\textsf{Channel Estimation Error (Self-Interference)}} + 
    \underbrace{\alpha \sum_{\text{k}\neq k} \sqrt{\frac{G_{k}P_{\text{k}}}{N_{\mathsf{a}}}} \boldsymbol{h}^*_{k}\boldsymbol{f}_{\text{k}}s_{\text{k}}}_{\textsf{Intra-Cell Interference}} +     \underbrace{ \sum_{\ell}\sum_{\text{k}=0}^{K^d_\ell-1} \sqrt{\frac{G_{\ell,k}P_{\ell,\text{k}}}{N_{\mathsf{a}}}}\boldsymbol{h}^*_{\ell,k} \boldsymbol{q}_{\ell}}_{\textsf{Aggregate AQNM}} \\&+ \underbrace{\alpha \sum_{\ell\neq 0}\sum_{\text{k}=0}^{K_{\ell}^d-1} \sqrt{\frac{G_{\ell,k}P_{\ell,\text{k}}}{N_{\mathsf{a}}}} \boldsymbol{h}^*_{\ell,k}\boldsymbol{f}_{\ell,\text{k}}s_{\ell,\text{k}}}_{\textsf{Inter-Cell Interference}} + \underbrace{\sum\limits_{\text{k}\neq k} \sqrt{T_{\text{k},k}P_{\ell,\text{k}}} g_{\text{k},k} s_{\text{k},u}}_{\textsf{Same Cell Inter-User Interference}}+ \underbrace{\sum_{\ell \neq 0}\sum\limits_{\text{k}=0}^{K_{\ell}^u-1} \sqrt{T_{(\ell,\text{k}),k}P_{\ell,\text{k}}} g_{(\ell,\text{k}),k} s_{\ell,\text{k},u}}_{\textsf{Other Cells Inter-User Interference}}  
 +  \underbrace{v_k}_{\mathsf{Noise}}
    \end{split}    
\end{equation}
\end{figure*}

\begin{theorem}\label{Theorem6}
For channel hardening and with matched filter precoder, the output SQINR of the $k$-th downlink user is expressed by (\ref{downlinksqinr}).
\end{theorem}
\begin{equation}\label{downlinksqinr}
\overline{\mathsf{sqinr}}_k^{\mathsf{MF}}  = \frac{ \alpha^2\frac{P_k}{P_u}\mathsf{SNR}_k^u \frac{P_k}{P_d}\mathsf{SNR}_k^d  N_{\mathsf{a}}}{\left( 1+\frac{P_k}{P_u}\mathsf{SNR}_k^u + \sum_{\ell \in \mathcal{C}}\frac{P_{\ell,k}}{P_u}\mathsf{SNR}_{\ell,k}^u \right) \overline{\mathsf{den}}^\mathsf{MF}}
\end{equation}
While introducing the new term $\varrho = \mathsf{SNR}_{\ell,(l,k)}^d/ \mathsf{SNR}_{\ell,(l,k)}^u$ for any $\ell, l$ and $k$ as the \textit{forward-reverse SNR ratio}, the output SQINR of the $k$-th downlink UE becomes 
\begin{equation}
\overline{\mathsf{sqinr}}^{\mathsf{MF}}_k = \frac{ \alpha^2 \frac{N_{\mathsf{a}}}{\varrho + \frac{P_k}{P_u} \mathsf{SNR}_k^d + \sum_{\ell \in \mathcal{C}} \frac{P_{\ell,k}}{P_u}\mathsf{SNR}_{\ell,k}^d  } \frac{P_{k}}{P_u} \frac{P_k}{P_d} \left(\mathsf{SNR}^d_k\right)^2 }{\overline{\mathsf{den}}^{\mathsf{MF}}}    
\end{equation}
where $\overline{\mathsf{den}}^{\mathsf{MF}}$ is given by (\ref{denTheorem6}), $\mathcal{C}$ is the set of cells reusing the same pilots and $\mathsf{SNR}^{\mathsf{iui}}_{(\ell,\text{k}),k} = T_{(\ell,\text{k}),k}P_u/N_0$.
\begin{figure*}
\begin{equation}\label{denTheorem6}
\begin{split}
\overline{\mathsf{den}}^{\mathsf{MF}} =&   1+ \alpha^2\sum_\ell \mathsf{SNR}_{\ell,k}^d + \alpha^2\sum_{\ell \in \mathcal{C}}  \frac{  N_{\mathsf{a}}}{\varrho+\frac{P_k}{P_u}\mathsf{SNR}_{\ell,k}^d + \sum_{l\in \mathcal{C}}\frac{P_{l,k}}{P_u}\mathsf{SNR}_{\ell,(l,k)}^d}\frac{P_{k}}{P_u}\frac{P_{\ell,k}}{P_d}\left(\mathsf{SNR}_{\ell,k}^d\right)^2  \\&+  \sum_{\ell}\sum_{\text{k}=0}^{K_\ell^u-1}\frac{P_{\ell,\text{k}}}{P_u}  \mathsf{SNR}_{(\ell,\text{k}),k}^{\text{iui}} + \alpha(1-\alpha) \sum_{\ell} \frac{P_{\ell,k}}{P_d}\mathsf{SNR}_{\ell,k}^d\left( K_\ell^d + 1 \right)  
\end{split}
\end{equation}
\end{figure*}
\begin{proof}
Referring to Corollary \ref{corchannelhardening}, \cite[Appendix A]{massiveadc} and after some mathematical manipulation, we retrieve the expression of the SQINR in Theorem \ref{Theorem6}. Note that by decomposing the inter-cell interference, the pilot contamination term can be extracted separately for the purpose of analysis. 
\end{proof}
\begin{remark}
The fractions $\frac{P_{\ell,k}}{P_u}$ and $\frac{P_{\ell,k}}{P_d}$ for any $\ell$ and $k$, are nothing but the power control and the power allocation coefficients, respectively.
\end{remark}
With $\overline{\mathsf{sqinr}}_k,~k=0,\ldots K-1$ are locally stable, the evaluation of the gross spectral efficiencies do not require averaging over the fading realizations, but rather it is directly computed as
\begin{equation}
    \frac{\bar{\mathcal{I}}_k}{B} = \log\left( 1 + \overline{\mathsf{sqinr}}_k \right),~k=0,\ldots,K-1
\end{equation}
If we incorporate the pilot overhead, this calls for an aggregation of the reverse and forward spectral efficiencies or, in its place, for a partition of the overhead between the reverse and forward links. The effective forward link spectral efficiency becomes
\begin{equation}
    \frac{\bar{\mathcal{I}}_k^{\mathsf{eff}}}{B} = \left(1 - \beta \frac{N_{\mathsf{p}}}{N_{\mathsf{c}}} \right) \log\left( 1 + \overline{\mathsf{sqinr}}_k \right),~k=0,\ldots,K-1
\end{equation}
where $\beta \in [0,1]$ is the fraction of the pilot overhead ascribed to the forward link, $N_{\mathsf{p}}$ is the number of pilots per cell and $N_{\mathsf{c}}$ is the fading coherence tile.

\begin{proposition}\label{prop5}
Considering a single-cell multiuser system (without any inter-cell interference) with perfect CSI, Corollary \ref{corchannelhardening} entails the results for forward link in \cite{massiveadc}.
\end{proposition}

\begin{proposition}\label{prop6}
With channel hardening, without full-duplexing (no inter-user interference), with full-resolution and matched filter precoder, the output SINR of the $k$-th downlink user is given by (\ref{downlinkhardeningsinr}).
\end{proposition}
\begin{figure*}
\begin{equation}\label{downlinkhardeningsinr}
\overline{\mathsf{sinr}}_k^\mathsf{MF} =  \frac{  \frac{N_{\mathsf{a}}}{\varrho + \frac{P_k}{P_u} \mathsf{SNR}_k^d + \sum_{\ell \in \mathcal{C}} \frac{P_{\ell,k}}{P_u}\mathsf{SNR}_{\ell,k}^d  } \frac{P_{k}}{P_u} \frac{P_k}{P_d} \left(\mathsf{SNR}^d_k\right)^2 }{ 1+ \sum_\ell \mathsf{SNR}_{\ell,k}^d + \sum_{\ell \in \mathcal{C}}  \frac{  N_{\mathsf{a}}}{\varrho+\frac{P_k}{P_u}\mathsf{SNR}_{\ell,k}^d + \sum_{l\in \mathcal{C}}\frac{P_{l,k}}{P_u}\textsf{SNR}_{\ell,(l,k)}^d}\frac{P_{k}}{P_u}\frac{P_{\ell,k}}{P_d}\left(\mathsf{SNR}_{\ell,k}^d\right)^2    }   
\end{equation}
\vspace*{-.5cm}
\end{figure*}
\begin{remark}
Note that Proposition \ref{prop6} entails the same result for forward link derived in \cite{massiveforward}.    
\end{remark}

Neglecting the pilot contamination, the output SINR of the $k$-th downlink user (\ref{downlinkhardeningsinr}) is reduced to (\ref{downlinksinr}). 

\begin{equation}\label{downlinksinr}
 \overline{\mathsf{sinr}}_k^\mathsf{MF} \approx \frac{\frac{P_{k}}{P_u} \frac{P_k}{P_d} \left(\mathsf{SNR}^d_k\right)^2N_{\mathsf{a}}}{\left(\varrho + \frac{P_k}{P_u} \mathsf{SNR}_k^d\right) \left( 1 + \sum_{\ell} \mathsf{SNR}_{\ell,k}^d  \right)     }   
\end{equation}

\section{Numerical Analysis}
In this section, we present the numerical results along with their discussion. The results are simulated with 10,000 Monte Carlo iterations. In the cell of interest, we evaluate the SQINR for each user with uniform power allocation and then we compute the average among all the users. Finally, we evaluate the CDF as well as the spectral efficiency for the average SQINR per cell. Unless otherwise stated, the values of the system parameters are defined in Table \ref{sysparam}.
\begin{table}[t]
\renewcommand{\arraystretch}{1}
\caption{System Parameters \cite{foundationsmimo,massiveforward}.}
\label{sysparam}
\centering
\begin{tabular}{rl}
%\hline
\textbf{Parameter} & \textbf{Value}\\
\hline
Bandwidth & 20 MHz\\
Pathloss Exponent ($\eta$) & 2.5\\
Shadowing ($\sigma_{\mathsf{dB}}$) & 8 dB \\
Downlink Transmit Power & 40 W\\
Uplink Transmit Power & 200 mW\\
Thermal Noise Spectral Density & -174 dBm/Hz\\
Noise Figure & 3 dB\\
BS Antennas Gain & 12 dB\\
Number of Antennas ($N_{\mathsf{a}}$) & 100\\
Uplink/Downlink Users per Cell ($K_\ell$) & 10 \\
Number of Pilots per Cell ($N_{\mathsf{p}}$) & 3$K_\ell$ \\
Fraction of Pilot Overhead ($\beta$) & 0.5\\
Fading Coherence Tile ($N_{\mathsf{c}}$) & 20,000 (Pedestrians)\\
ADC/DAC resolution & 3 bits
%\hline
\end{tabular}
\end{table}

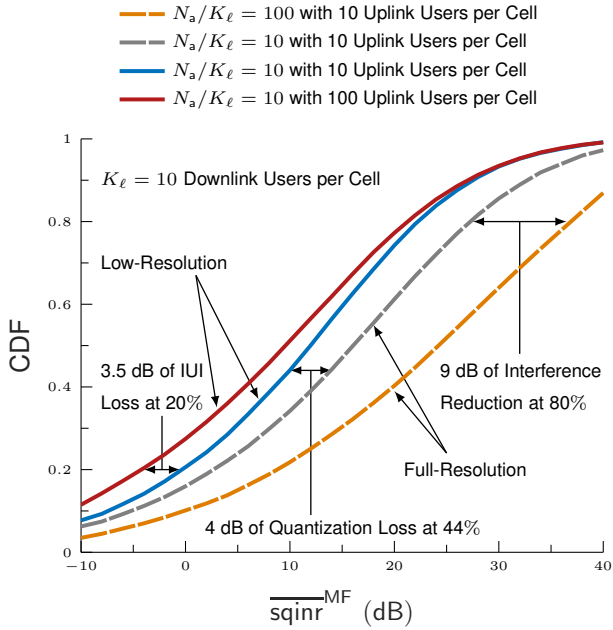
\begin{figure}[t]
\centering
\setlength\fheight{5.5cm}
\setlength\fwidth{7.3cm}
%\definecolor{mycolor1}{rgb}{0.00000,0.44700,0.74100}%
%\definecolor{mycolor2}{rgb}{0.85000,0.32500,0.09800}%
%\definecolor{mycolor3}{rgb}{0.92900,0.69400,0.12500}%
%\definecolor{mycolor4}{rgb}{0.49400,0.18400,0.55600}%

\definecolor{mycolor1}{rgb}{1.00000,0.00000,1.00000}%
\definecolor{mycolor2}{rgb}{0.00000,0.49804,0.00000}%
\definecolor{mycolor3}{rgb}{0.00000,0.44706,0.74118}%
\definecolor{mycolor4}{rgb}{0.87059,0.49020,0.00000}%
\definecolor{cornellred}{rgb}{0.7, 0.11, 0.11}
\begin{tikzpicture}

\begin{axis}[%
width=0.951\fwidth,
height=\fheight,
at={(0\fwidth,0\fheight)},
scale only axis,
xmin=-10,
xmax=40,
xlabel style={font=\color{white!15!black}},
xlabel={$\mathsf{\overline{sqinr}^{MF}~(dB)}$},
ymin=0,
ymax=1,
ylabel style={font=\color{white!15!black}},
ylabel={\textsf{CDF}},
axis background/.style={fill=white},
axis x line*=bottom,
axis y line*=left,
legend style={at={(.9,1.35)},legend cell align=left, align=left, draw=none}
]

\node[right, align=left, rotate=0]
at (axis cs:-9,.7) {\scriptsize\sffamily{Low-Resolution}};
\draw [-latex,black,line width=.5pt] (1,.67) to (3,.35);
\draw [-latex,black,line width=.5pt] (1,.67) to (7,.37);

\node[right, align=left, rotate=0]
at (axis cs:20,.2) {\scriptsize\sffamily{Full-Resolution}};
\draw [-latex,black,line width=.5pt] (25,.24) to (20,.39);
\draw [-latex,black,line width=.5pt] (25,.24) to (18,.55);

\node[right, align=left, rotate=0]
at (axis cs:1,.06) {\scriptsize\textsf{4 dB of Quantization Loss at 44$\%$}};
\draw [-,black,line width=.5pt] (12,.44) to (12,.09);
\draw [latex-latex,black,line width=.5pt] (10,.44) to (14,.44);

\node[right, align=left, rotate=0]
at (axis cs:-9,.4) {\scriptsize\textsf{3.5 dB of IUI }\vspace*{-5cm}\\\scriptsize\textsf{Loss at 20$\%$}};

\draw [latex-latex,black,line width=.5pt] (-4,.2) to (-.5,.2);
\draw [-,black,line width=.5pt] (-2.25,.2) to (-2.25,.33);

\node[right, align=left, rotate=0]
at (axis cs:-9,.9) {\scriptsize$K_\ell = 10~$\textsf{Downlink Users per Cell}};

\node[right, align=left, rotate=0]
at (axis cs:23.5,.4) {\scriptsize\textsf{9 dB of Interference}\\\scriptsize\textsf{Reduction at 80$\%$}};
\draw [latex-latex,black,line width=.5pt] (27.5,.8) to (36.5,.8);
\draw [-,black,line width=.5pt] (32,.8) to (32,.48);

\addplot [color=mycolor4,dash pattern={on 10pt off 1pt on 0pt off 0pt},line width=1.5pt]
  table[row sep=crcr]{%
-10	0.0349\\
-8	0.0447666666666667\\
-6	0.05732\\
-4	0.06982\\
-2	0.08446\\
0	0.1013\\
2	0.11812\\
4	0.13792\\
6	0.1628\\
8	0.18864\\
10	0.21798\\
12	0.25064\\
14	0.28542\\
16	0.32168\\
18	0.36134\\
20	0.40284\\
22	0.44738\\
24	0.49468\\
26	0.54282\\
28	0.59156\\
30	0.6399\\
32	0.6877\\
34	0.73438\\
36	0.7795\\
38	0.824533333333333\\
40	0.8694\\
};
%\addlegendentry{Na/K = 100, Full-Resolution, 10 Uplink Users per Cell}
\addlegendentry{\scriptsize{\sffamily{$N_{\mathsf{a}}/K_\ell = 100$ with 10 Uplink Users per Cell}}}
\addplot [color=gray,dash pattern={on 10pt off 1pt on 0pt off 0pt}, line width=1.5pt]
  table[row sep=crcr]{%
-10	0.0625\\
-8	0.0744666666666667\\
-6	0.09244\\
-4	0.1118\\
-2	0.1337\\
0	0.1597\\
2	0.18922\\
4	0.2214\\
6	0.2566\\
8	0.2984\\
10	0.34186\\
12	0.39048\\
14	0.44226\\
16	0.49866\\
18	0.55466\\
20	0.61212\\
22	0.6684\\
24	0.72064\\
26	0.77114\\
28	0.81616\\
30	0.85596\\
32	0.88904\\
34	0.91894\\
36	0.93954\\
38	0.959866666666667\\
40	0.9728\\
};
%\addlegendentry{Na/K = 10, Full-Resolution, 10 Uplink Users per Cell}
\addlegendentry{\scriptsize{\sffamily{$N_{\mathsf{a}}/K_\ell = 10$ with 10 Uplink Users per Cell}}}

\addplot [color=mycolor3, line width=1.5pt]
  table[row sep=crcr]{%
-10	0.0772\\
-8	0.0930666666666667\\
-6	0.1166\\
-4	0.14124\\
-2	0.1715\\
0	0.2059\\
2	0.24172\\
4	0.28458\\
6	0.33424\\
8	0.38684\\
10	0.44038\\
12	0.50102\\
14	0.5644\\
16	0.62574\\
18	0.68482\\
20	0.7424\\
22	0.79442\\
24	0.8386\\
26	0.87574\\
28	0.90746\\
30	0.93318\\
32	0.95204\\
34	0.96604\\
36	0.97624\\
38	0.9849\\
40	0.9921\\
};
\addlegendentry{\scriptsize{\sffamily{$N_{\mathsf{a}}/K_\ell = 10$ with 10 Uplink Users per Cell}}}

\addplot [color=cornellred,line width=1.5pt]
  table[row sep=crcr]{%
-10	0.115\\
-8	0.142766666666667\\
-6	0.17264\\
-4	0.20344\\
-2	0.23658\\
0	0.2744\\
2	0.3153\\
4	0.35988\\
6	0.40772\\
8	0.45782\\
10	0.5117\\
12	0.56574\\
14	0.61886\\
16	0.67328\\
18	0.72566\\
20	0.7726\\
22	0.81544\\
24	0.85384\\
26	0.8857\\
28	0.91292\\
30	0.93486\\
32	0.9529\\
34	0.96728\\
36	0.97764\\
38	0.985933333333333\\
40	0.9914\\
};
%\addlegendentry{Na/K = 10, Low-Resolution, 100 Uplink Users per Cell}
\addlegendentry{\scriptsize{\sffamily{$N_{\mathsf{a}}/K_\ell = 10$ with 100 Uplink Users per Cell}}}

\end{axis}
\end{tikzpicture}%
    \caption{Forward link results: Effects of full-duplexing, ADC/DAC quantization error, and the factor $N_{\mathsf{a}}/K_\ell$ on the CDF of the SQINR. The dashed gray and solid blue curves are simulated for full and low resolution DAC, respectively.}
    \label{fig1}
\end{figure}
Fig.~\ref{fig1} plots the CDF or the outage probability of the average forward link SQINR. When the number of uplink users is large (100 users per cell) and hence large IUI, the performance gets worse and vice versa. In addition, the performance improves with increasing the ADC/DAC resolution.
Moreover, by largely increasing the ratio $N_{\mathsf{a}}/K_\ell$, the user beams exceedingly sharp, the desired signal dominates over the noise and the interference. This factor for example, rejects about 9 dB of interference at a CDF of 80$\%$.

\begin{figure}[t]
\centering
\setlength\fheight{5.5cm}
\setlength\fwidth{7.3cm}
\input{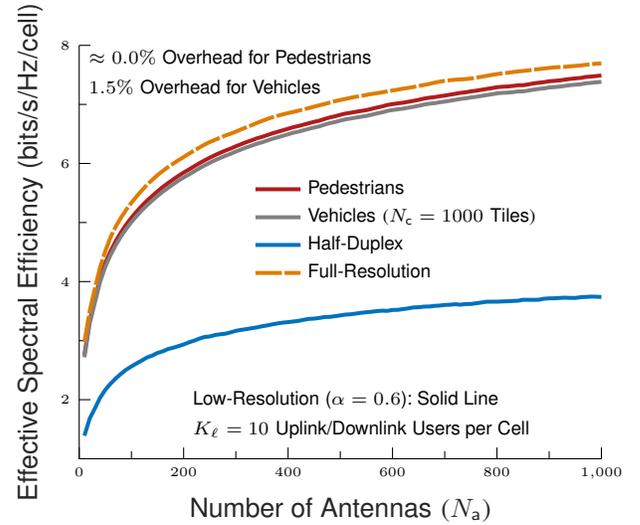}
    \caption{Forward link results: Effects of SI power, overhead, DAC resolution, duplexing mode and pilot contamination on the spectral efficiency. Dashed yellow curve stands for full-resolution, pedestrians and full-duplex. Solid red and gray curves stand for low-resolution full-duplex. Solid blue curve stands for low-resolution DAC, half-duplex and pedestrians scenario.}
    \label{fig2}
\end{figure}
Fig.~\ref{fig2} shows the forward link spectral efficiency increasing with the number of antennas at the BS and quantization bits. Since the pedestrians feature large fading coherence tile and hence low overhead, they achieve better spectral efficiency compared to the vehicles (high mobility) feature small fading coherence tile and hence higher overhead. Besides, we observe that the proposed system outperforms half-duplex mode, which shows the feasibility of FD in cellular networks.

\section{Conclusion}
In this work, we provide a unified framework for forward link FD massive MIMO cellular networks with low-resolution ADCs/DACs and under pilot contamination. Using matched filter precoder at the BS, AQNM modeling for ADCs and DACs, and channel hardening, we analyzed the SQINR CDF and spectral efficiency for hexagonal lattice with two tiers. Using a proper scaling ratio of antennas over the number of users per cell rejects large amount of interference, however, the SINR is limited by the pilot contamination. Simulation results show that the quantization error as well as the IUI (caused by the FD BS) incur losses; however, this loss is compensated by employing a massive number of antennas. Finally, the proposed system outperforms half-duplex mode in terms of spectral efficiency which is part of the goal of this work to show the feasibility of FD in cellular networks.

\balance
\bibliographystyle{IEEEtran}
\bibliography{main}
\end{document}